\documentstyle[psfig]{mn}

\title[ROSAT PSPC detection of soft X-ray absorption in GB~1428+4217]
{ROSAT PSPC detection of soft X-ray absorption in GB~1428+4217:
The most distant matter yet probed with X-ray spectroscopy}

\author[Th. Boller, A.C. Fabian, W.N. Brandt and M.J. Freyberg]
{Th. Boller$^1$, A.C. Fabian$^2$, W.N. Brandt$^3$ and M.J. Freyberg$^1$  \\
$^1$Max-Planck-Institut f\"ur extraterrestrische Physik,
Postfach 1603, 85748 Garching, Germany \\
$^2$Institute of Astronomy, Madingley Road, Cambridge CB3 0HA\\
$^3$Department of Astronomy and Astrophysics, Penn State,
525 Davey Lab, University Park, PA 16802, USA
\\
}
\date{Received 2000 February 20; Accepted 2000 May 4}

\begin{document}
\label{firstpage}

\maketitle

\begin{abstract}
We report on a ROSAT PSPC observation of the highly-luminous $z = 4.72$ 
radio-loud quasar GB~1428+4217 obtained between 1998 December 11 and 17,
the final days of the ROSAT satellite.
The low-energy sensitivity of the PSPC detector was employed to constrain
the intrinsic X-ray absorption of the currently most distant X-ray detected
object. Here we present the detection of significant soft X-ray absorption   
towards GB~1428+4217, making the absorbing material the most distant matter 
yet probed with X-ray spectroscopy. X-ray variability by 
$25\pm 8$ per cent is detected on a timescale of 6500 s in the rest
frame.
The X-ray variation requires an unusually high 
radiative efficiency $\eta$ of at least 4.2, further supporting the blazar 
nature of the source. 
\end{abstract}


\section{Introduction}
Observations of high redshift quasars are of wide cosmological 
importance since these objects are thought to be associated with the earliest
collapsed structures. Many high-redshift radio-loud quasars have recently
been found to show low-energy X-ray cutoffs, and these are believed to
be associated with intrinsic X-ray absorbers of column densities several times
$\rm 10^{22}\ cm^{-2}$ (e.g. 
Elvis et al. 1994a;
%Cappi et al. 1997;
Elvis et al. 1998; 
Fiore et al. 1998).
The radio-loud quasar GB~1428+4217 is currently the most distant X-ray
detected object. It has been studied with the ROSAT HRI and ASCA
(Fabian et al. 1997, 1998). These observations revealed an extreme isotropic
X-ray luminosity of about $\rm 1.3 \times 10^{47}\ erg\ s^{-1}$ as well
as variability by a factor of 2 on a timescale of about 2.4 days in the 
rest frame.

The ASCA observations did not allow tight constraints 
to be placed on 
X-ray absorption due to the lack of low-energy response in the 0.1--0.5 keV
band and the limited low-energy calibration.
The ROSAT HRI did not give the needed spectral
information. Therefore, we proposed GB~1428+4217 for observation during the
last ROSAT observations in December 1998, employing the excellent
low-energy sensitivity of the PSPC detector to constrain the intrinsic
X-ray absorption in GB~1428+4217 as well as search for flux and
associated spectral variability. 

A value of the Hubble constant of $H_0$=$\rm 70\ km\ s^{-1}\ Mpc^{-1}$
and a cosmological deceleration parameter of $q_0 = \rm \frac{1}{2}$ have
been adopted throughout.

\section{The ROSAT PSPC observations from December 1998}
\subsection{Data quality assessment}
The ROSAT PSPC observations of GB~1428+4217 were performed in
the final observation period which was affected by some anomalies. 
Therefore careful data reduction and cleaning were required.
The first anomaly was an occasional breakdown of the high voltage
which was not recorded by the housekeeping system. These periods can be
identified by unusually low master veto rates ($ MV < 35\,\mbox{counts\,s}^{-1}$).
Moreover, the high voltage showed occasional flickering which caused
periods of frequent short accepted time intervals ($< 10$\,s) where
the nominal voltage (and thus gain value) may not be reached.
Finally, a gain hole in the PSPC detector began to develop which can 
currently not be corrected for and came close to the on-axis position 
after 1998 December 13 (cf. the December month page of the ROSAT 2000
calendar). 
We therefore selected only time intervals before 
1998 December 13 with master
veto rates between 35 and 300\,counts\,s$^{-1}$ and durations exceeding 100\,s
(see summary in Table \ref{tab_180308p}). 
We have verified that within our time selection the on-axis position is
not affected by the presence of the gain hole 
employing the nominal gain-corrected energy image of the Al K$\alpha$ line.
\begin{table}
\caption[]{
Observation log of 180308p (selected period)
}
\label{tab_180308p}
\begin{flushleft}
\begin{tabular}{ll}\hline
Start UTC  & 1998-Dec-11 17:17        \\
Start SSC  & 269122244                \\
End UTC    & 1998-Dec-12 09:35        \\
End SCC    & 269180932                \\
Exposure   & 9446\,s                 \\
Average MV & 114.9 \,counts\,s$^{-1}$    \\
Eff.gain   & 102.7                    \\
\end{tabular}
\end{flushleft}
\end{table}

Additionally, we wanted to ensure that residual gain uncertainties 
do not affect the data quality. We have checked the onboard Al\,$K\alpha$
measurements that were taken at 
spacecraft clock times (approximate seconds
after launch)
$SCC = 269147793$--$269148181$ (1998-Dec-12, 00:23 -- 00:29 UT)
and at $SCC = 269196947$--$269197310$ 
(1998-Dec-12, 14:02 -- 14:08 UT)
and could not find significant variations of the peak of the pulse height 
distribution. Moreover, effects due to gain uncertainties dominate at
higher pulse height amplitudes and our analysis concentrates on the
softer energy range. Nevertheless, we re-processed our events with varying
gain values ($\pm 1.4$ channels) to confirm that this effect is 
unimportant in our case (see the Appendix).

We note that calibration uncertainties in the very softest
channels may be present, and some authors find indications that the
ROSAT PSPC effective area may actually be larger 
than expected (e.g. Wolff et al. 1996). This would cause the column densities
measured below to be systematically low, and it would only increase
the intrinsic absorption in GB~1428+4217.

\subsection{X-ray data analysis}
The analysis of the cleaned ROSAT PSPC data 
was performed using the EXSAS software package (Zimmermann et al. 1994).   
The centroid position of GB~1428+4217 in the ROSAT PSPC image is
R.A.(2000) = $\rm 14^h30^m22.0^s \pm 0.72^s$,
Dec.(2000) = $\rm 42^{o}04^{'}41^{''} \pm 11^{''}$. The internal PSPC
position error is about 20 arcsec.
The total exposure spent on the source was 9446 seconds.
The source counts were obtained 
using a circular source cell of radius 
2.33 arcmin. The number of source plus background
counts within that cell is 1669 $\pm$ 41.
The background was determined from a source-free circular cell with
a radius of 8.1 arcmin
centered at
R.A.(2000) = $\rm 14^h28^m59.7^s$, 
Dec.(2000) = $\rm 42^{o}09^{'}41^{''}$
(a nearby X-ray source at a distance of 4.5 arcmin 
prevented us from using an annulus around the 
centroid position of GB~1428+4217).

The number of background counts expected in the source cell is 190 $\pm$ 14.
The net source counts are therefore 1479 $\pm$ 43, resulting
in a mean count rate of 0.156 $\pm$ 0.004 $\rm counts\ s^{-1}$
in the 0.1--2.4 keV energy range. The timing properties of GB~1428+4217
are discussed in Section 4.

\begin{figure}
       \psfig{figure=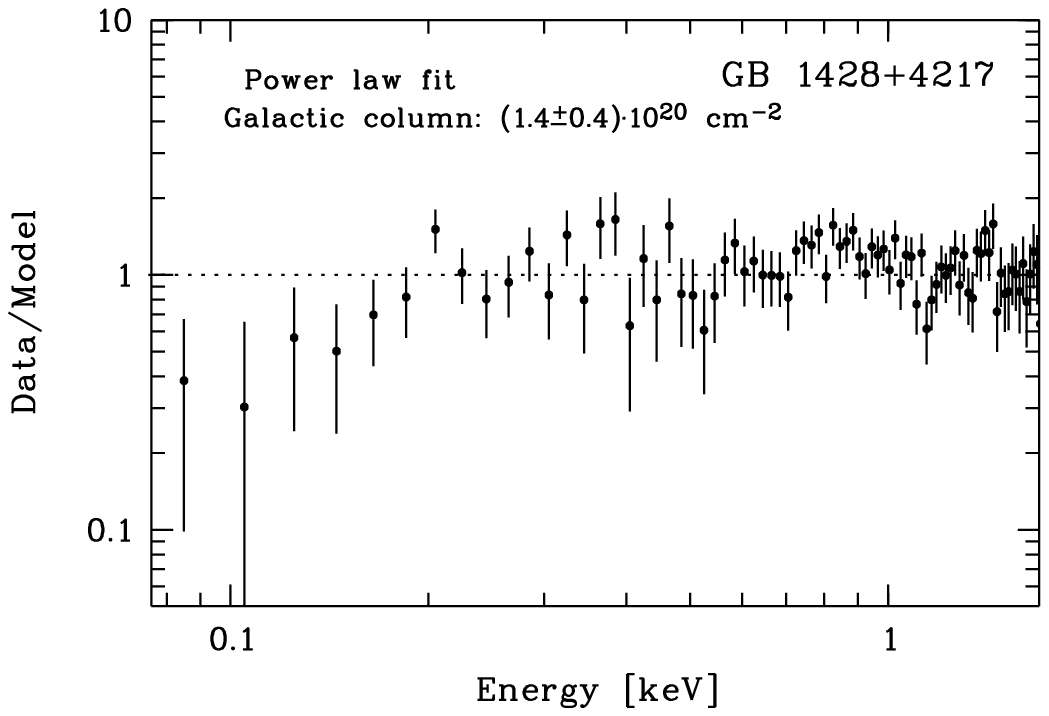,width=8.3cm,clip=}
       \psfig{figure=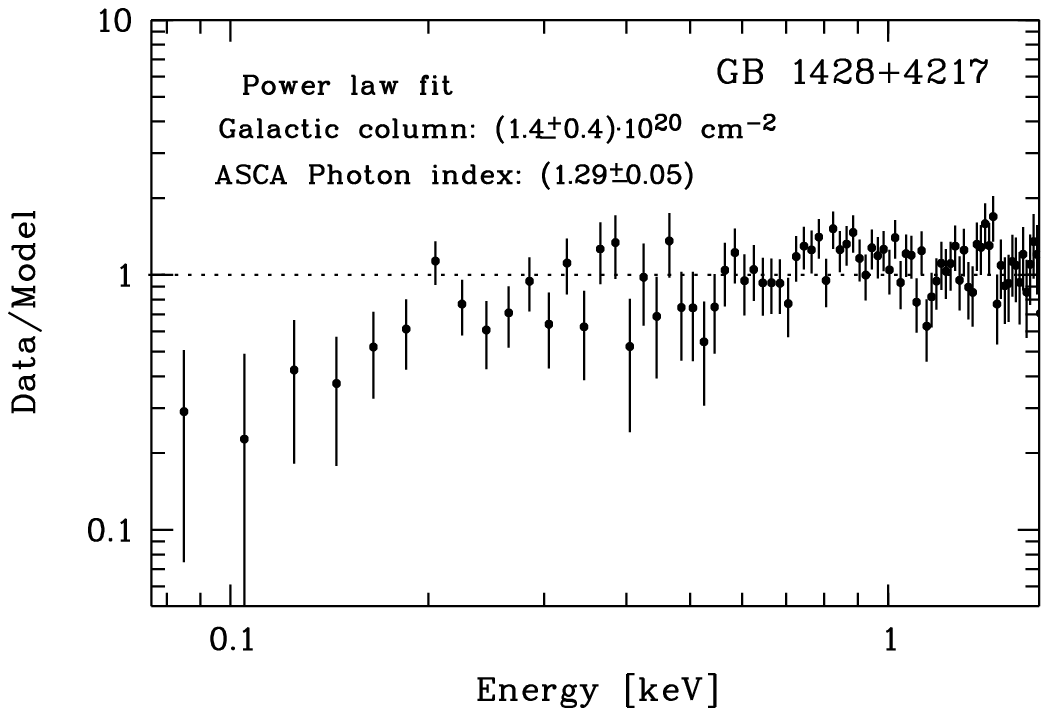,width=8.3cm,clip=}
      \caption{
{\bf Upper panel:} 
Power-law plus Galactic column fit 
to the ROSAT PSPC observations obtained
in December 1998 of the $z = 4.72$ radio-loud quasar GB~1428+4217. 
The column density was allowed to vary within its measurement error 
$N_{\rm H, Gal} \rm = (1.4 \pm 0.4) \times 10^{20}\ cm^{-2}$. 
While this fit is statistically acceptable, 
it shows clear systematic residuals at low energies,
and the derived photon index of 1.04$\pm$0.18 
is substantially flatter than that measured by ASCA (1.29$\pm$ 0.05; 
Fabian et al. 1998).
{\bf Lower panel:} The photon index was allowed to vary within the
range found by ASCA. Such a model can be 
statistically rejected with a probability of 98.7 per cent using the
$\chi^2$ distribution.
}
\end{figure}

\subsection{Constraining the intrinsic X-ray absorption}
\begin{figure}
       \psfig{figure=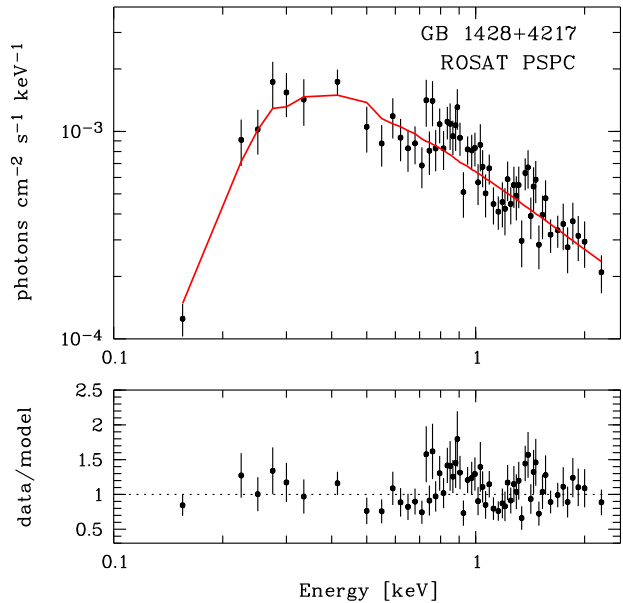,width=8.3cm,clip=}
      \caption{Power-law fit to the ROSAT PSPC observations obtained
in December 1998 of the $z = 4.72$ radio-loud quasar GB~1428+4217. In the
observer's frame the neutral absorbing column density 
is 
$N_{\rm H, fit}^{\rm z=0} \rm = (3.14 \pm 0.35) \times 10^{20}\ cm^{-2}$.
Assuming intrinsic X-ray absorption the absorbing column is
$N_{\rm H, fit}^{\rm z=4.72} \rm = (1.52 \pm 0.28) \times 10^{22}\ cm^{-2}$.
}
\end{figure}
\begin{figure}
       \psfig{figure=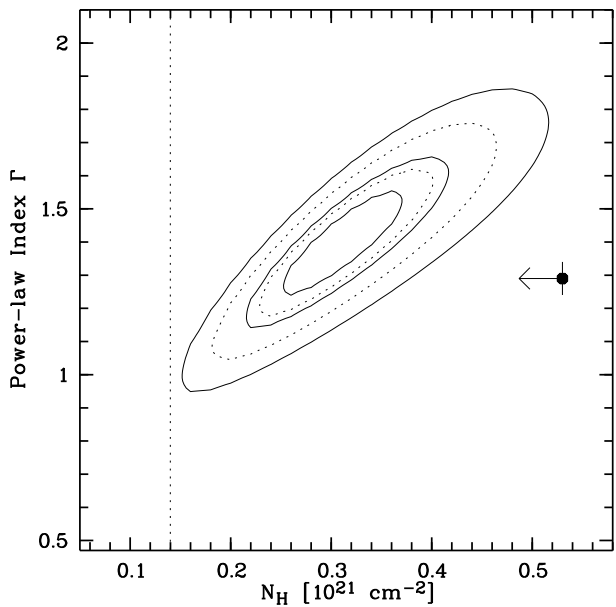,width=8.3cm,clip=}
      \caption{Contour plot of $\rm \chi^2$ as a function of the 
absorbing column $N_{\rm H, fit}^{\rm z=0}$ (x-axis) and the photon index 
$\rm \Gamma$ (y-axis) for five confidence levels of
68.3, 95.4, 99.7, 99.99 and 99.9999 per cent. The absorbing column is given in
units of $\rm 10^{21}\ cm^{-2}$. 
The Galactic column (dashed vertical line)
is $N_{\rm H, Gal} \rm = 1.4 \times 10^{20}\ cm^{-2}$.
The 90 per cent confidence upper limit to the
column density of $\rm 5.3 \times 10^{20}\ cm^{-2}$
and the corresponding ASCA photon index of $\rm \Gamma = 1.29 \pm 0.05$
obtained by Fabian et al. (1998) 
is marked by the left arrow.
}
\end{figure}

In the following we constrain the spectral properties of GB~1428+4217. 
The confidence intervals for the ROSAT fit parameters presented below
correspond to 90 per cent intervals.
First, we have performed a power-law fit 
(cf. Figure~1) with the 
neutral absorption column density constrained to be
consistent with the Galactic value of
$N_{\rm H, Gal}=(1.4\pm 0.4)\times 10^{20}$~cm$^{-2}$
 (Dickey \& Lockman 1990; Elvis et~al. 1994b).
 While this fit is statistically acceptable ($\chi^2$=120 for 114
  d.o.f.), it shows clear systematic residuals at low energies 
  (upper panel of Figure 1), and the derived photon index of 1.04$\pm$0.18 
  is substantially flatter than that measured by ASCA (1.29$\pm$0.05; 
  Fabian et al. 1998). While we cannot statistically rule out a 
  flattening of the intrinsic continuum at low energies, such flattenings
  are not usually seen in active galaxies, and the intrinsic continuum
  would need to be exceptionally hard. 
  Furthermore, we note that
  ROSAT versus ASCA calibration discrepancies cannot easily explain a
  systematically flatter ROSAT continuum (ROSAT slopes, if anything, 
  appear to be systematically steeper than ASCA slopes; see 
  Iwasawa, Fabian \& Nandra 1999). If we require the photon index
  to lie within the ASCA range of 1.29$\pm$0.05 and the column 
  density to lie within the range consistent with the Galactic
  value, the fit can be statistically rejected with a probability 
  of 98.7 per cent ($\chi^2$=150 for 114 d.o.f.; lower
  panel of Figure 1). We conclude that plausible power-law models 
  absorbed by the Galactic column density cannot provide an adequate 
  fit to the ROSAT data.

A simple power-law fit
where the absorption column density and the photon index are
allowed to be free parameters
provides a better statistical fit
to the ROSAT PSPC data ($\chi^2$ = 100 for 113 d.o.f.; cf. Figure 2).
Using the $F$-test for the addition of one free parameter one gets
$\Delta \chi^2/\chi_{\nu}^2$ = 22.7 (cf. Eq. 11.50 of
Bevington \& Robinson 1992).  According to Table C.5 of Bevington \& Robinson
(1992) this corresponds to a highly significant improvement
($>$ 99.9 per cent) in the 
fit quality.
Most important, the soft X-ray
absorption of $N_{\rm H, fit}^{\rm z=0}\rm=(3.14\pm0.35)\times10^{20}\ cm^{-2}$
is larger than the  Galactic column towards GB~1428+4217 at the 5 sigma level.
The photon index is $\Gamma$=1.40$\pm$0.20, in agreement
with the ASCA value.
These results are robust to changes in the number of data points included
in the fit. A contour plot of $\rm \chi^2$ as a function of the
absorbing column in the observer's frame and the photon index is shown
in Figure 3. The displayed contour levels of 68.3, 95.4, 
99.7, 99.99 and 99.9999 per cent clearly demonstrate the detection of excess absorption 
above the Galactic column towards the high redshift quasar
GB~1428+4217.
This excess low-energy cutoff translates into an intrinsic
X-ray absorption of $N_{\rm H, fit}^{\rm z=4.72} \rm 
= (1.52 \pm 0.28) \times 10^{22}\ 
cm^{-2}$ for neutral gas and solar abundances.
With the current data we are unable to precisely constrain the ionization 
level of the absorbing gas; our only constraint is that oxygen must not 
be completely stripped. However, we note that increasing the gas'
ionization level will require even larger column densities than the
substantial one already derived for neutral gas.

In the Appendix we have modeled the influence of different gain
values on the soft X-ray absorption. With Figure A1 in the Appendix 
we demonstrate that even in the case of unusual gain values the 
excess absorption above
the Galactic column is still present.

\begin{figure}
       \psfig{figure=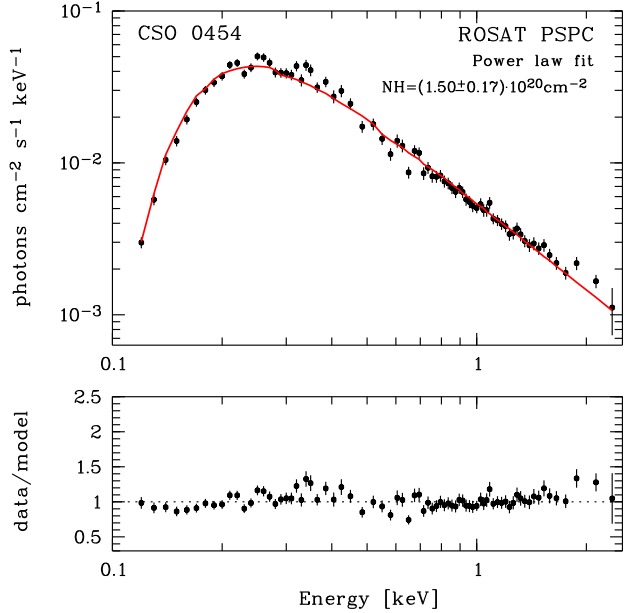,width=8.3cm,clip=}
      \caption{Power-law fit to the ROSAT PSPC observations obtained
in December 1998 of CSO 0454. 
The column density resulting from the fit is 
$N_{\rm H, fit} \rm = (1.50 \pm 0.17) \times 10^{20}\ cm^{-2}$. 
This value is significantly below $\rm (3.14 \pm 0.35) \times 10^{20}
cm^{-2}$ found for the GB~1428+4217 further supporting the presence
of significant soft X-ray absorption in GB~1428+4217.
}
\end{figure}
\begin{figure}
       \psfig{figure=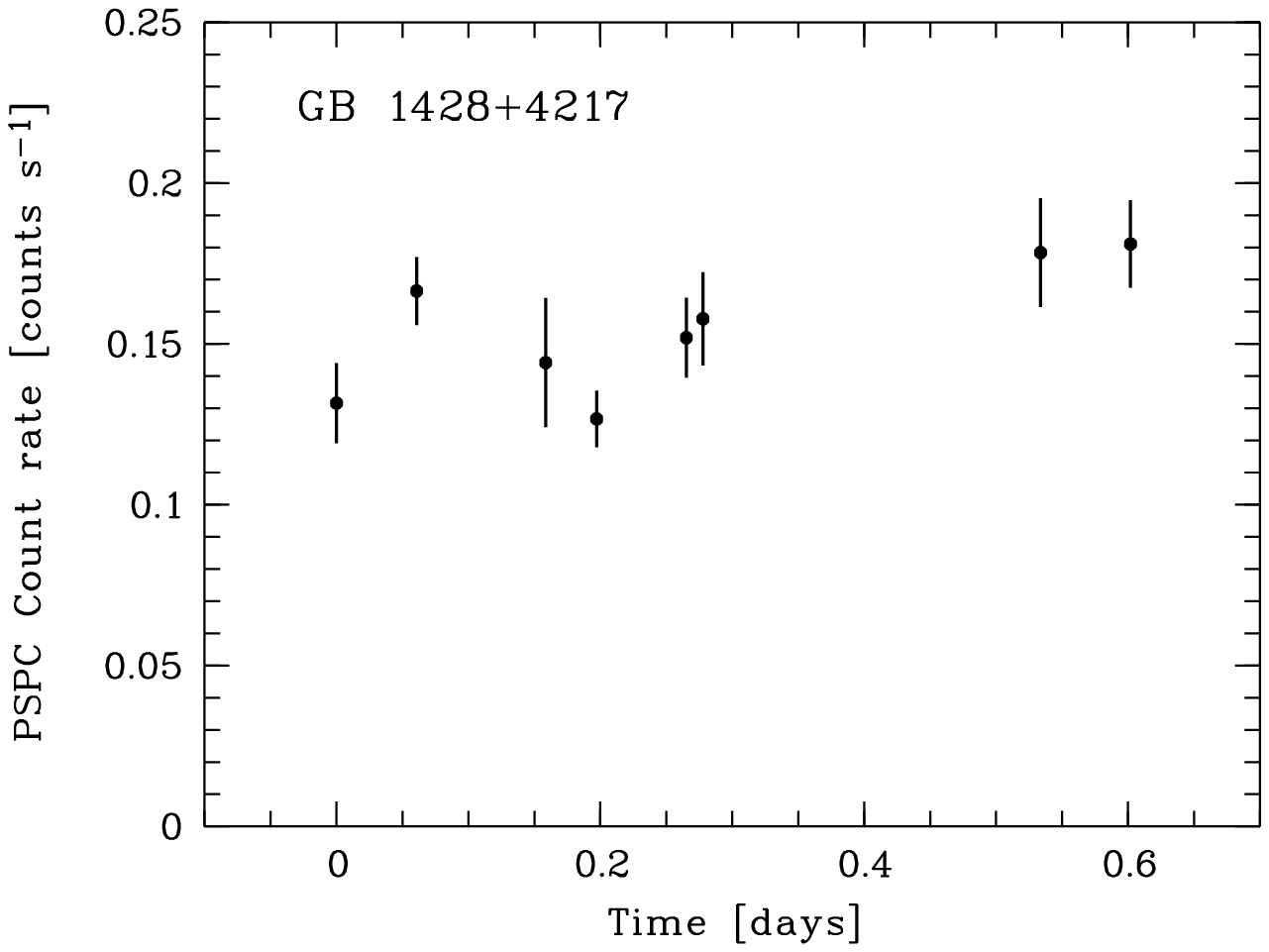,width=8.3cm,clip=}
       \psfig{figure=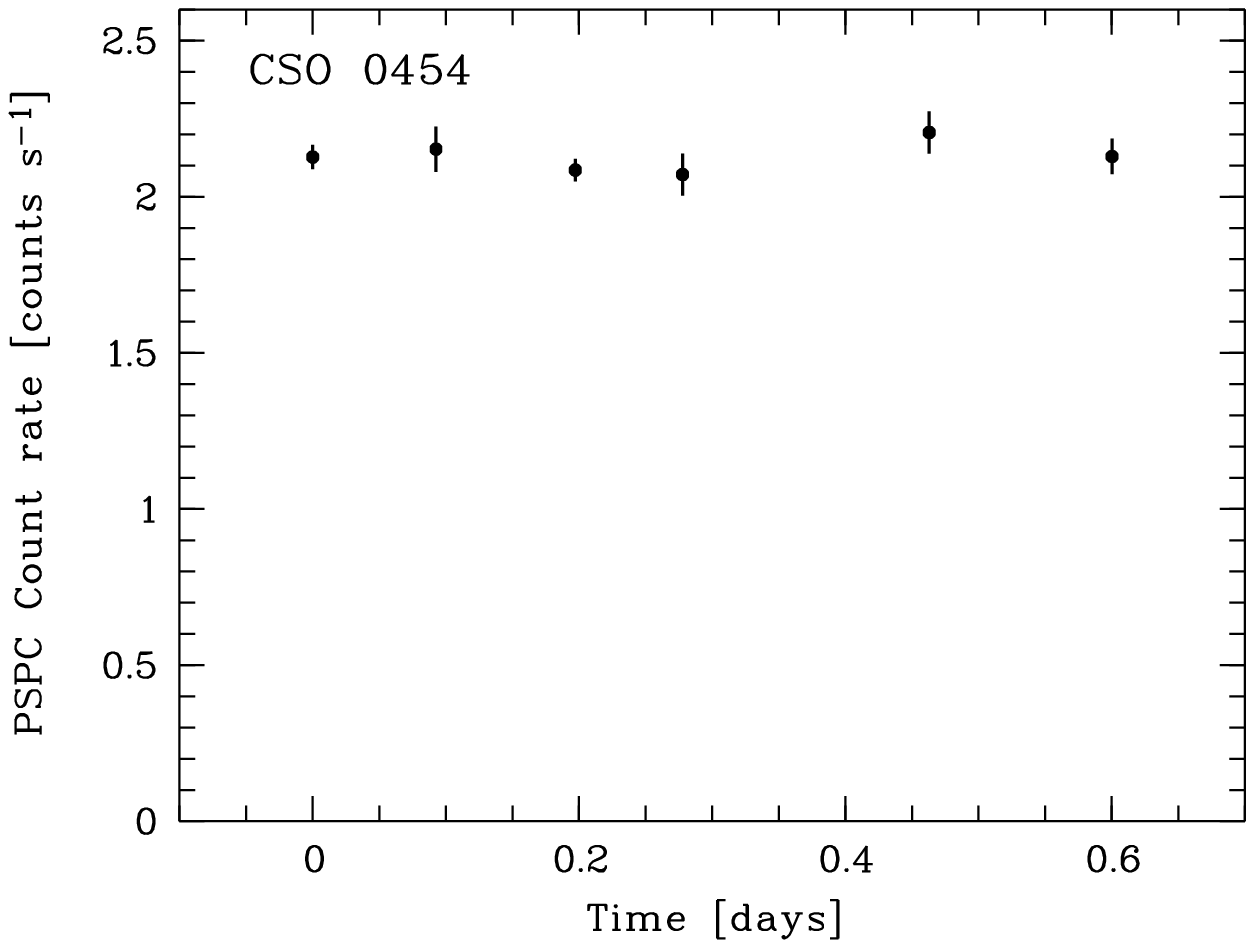,width=8.3cm,clip=}
      \caption{
{\bf Upper panel:}
ROSAT PSPC light curve of GB~1428+4217
obtained between 1998 December 11 and December 12.
Variability by about 25 $\pm$ 8 per cent on a timescale of about
37000 s is detected in the observer's frame.
{\bf Lower panel:}
ROSAT PSPC light curve of CSO 0454. The count rate
variations detected in the observer's frame in GB~1428+4217 
do not occur in CSO 0454. 
}
\end{figure}

%\begin{figure}
%       \psfig{figure=gb1428_light_98.ps,width=8.3cm,clip=}
%      \caption{
%ROSAT PSPC light curve of GB~1428+4217 
%obtained between 1998 December 11 and December 12.
%Amplitude variability of about 25 $\pm$ 8 per cent on a timescale of about
%37000 s are detected in the observer's frame. 
%}
%\end{figure}

Interestingly, there is another bright X-ray source 
(the blazar CSO 0454 = 1H 1430+423, $z=0.129$) 
in the field
of view, which can be used to demonstrate that there is not a relevant
systematic effect in determining the excess absorption. 
The source is outside the gain hole and no other effects prevented the
reliable analysis of the data for this source.
A power-law fit combined with absorption by neutral matter (Figure 4) results
in $N_{\rm H, fit}^{\rm z=0} \rm = (1.50 \pm 0.17) \times 10^{20}\ cm^{-2}$, 
in good agreement with the Galactic column density.
In addition, the X-ray variability found in GB~1428+4217
does not occur in CSO 0454 (see Section 4 and Figure 5).

%We have also fitted the ROSAT PSPC data by varying the 
%number densities of the element abundances in the Galactic column.
%To obtain an absorption value consistent with the Galactic column,
%a helium overabundance of a factor of 2.6 is required,
%26 per cent helium instead of 10 per cent as used in the fitting described above.
%The ROSAT data do not allow discrimination between excess absorption
%or a helium overabundance. 
%Laor et al. (1997) find that helium overabundances are not usually
%seen in quasar spectra (see their Section 4.2). 
%Therefore, we do not
%consider a helium overabundance as the likely explanation for the
%excess low-energy cutoff.
%The ROSAT data rule out single abundance changes by
%elements `later' than carbon.
%For carbon an overabundance of about 10 is required to obtain absorption
%consistent with the Galactic column, and the spectral fit quality
%is significantly degraded.

GB~1428+4217 is clearly detected in the ROSAT All-Sky Survey with 
a detection likelihood of 35 (see Cruddace et al. 1988 for the
definition of the detection likelihood). The ROSAT PSPC count rate
is $\rm 0.046 \pm 0.011\ counts\ s^{-1}$. The exposure time is 639 seconds.
The number of source photons detected during the survey observations
do not allow precise spectral fitting and the soft X-ray absorption
cannot be well constrained.

The residuals of the power-law fit to GB~1428+4217 (cf. Figure 2) indicate
the presence of peculiar  `wiggles' between about 0.8 and 1.4 keV. 
Their  origin is presently unclear. The dip at 1.2 keV
is near the observed frame energy for the iron K$\alpha$ line. 
The spectral energy resolution of the
PSPC detector is not able to resolve any iron emission line shifted
into the ROSAT energy band. These features may therefore only be due to
noise. 
Neither the ASCA data (Fabian et al. 1998)
nor the ROSAT All-Sky Survey data can rule out the presence of intrinsic 
residuals at the level seen in the present dataset. 
XMM-Newton observations of GB~1428+4217 may help to solve the origin of these 
peculiarities.

\section{Origin of the excess absorption}

Miralda-Escude (2000) has discussed the possibility that absorption by
singly ionized helium (He~{\sc ii}) in the intergalactic medium (IGM) might
be detectable in the soft X-ray spectra of distant quasars. As
GB~1428+4217 is the most distant X-ray loud quasar known, we have studied
the effect of He~{\sc ii} absorption in some detail. The analytic fit to the
absorption cross-section presented by Verner et al. (1996) 
and values of $\Omega_{\rm b} = 0.06$ and $\Omega_{\rm 0} = 0.3$ were used.
Unfortunately the effect of He~{\sc ii} absorption is completely swamped by
that of Galactic absorption in the direction of GB~1428+4217 which almost
completely absorbs all 100~eV X-rays and considerably depletes them at
200~eV. If all the helium in the IGM is He~{\sc ii} down to redshift 3, then
it only causes a further 30 per cent drop in the observed flux at
200~eV which is quite insufficient to explain our observed spectrum
(see Figure~2). In order to use soft X-ray absorption to measure
intergalactic helium, a bright quasar, preferably a blazar, at $z>4$
in the Lockman Hole is required.

Another, speculative, possibility is that we have detected absorption by
intergalactic oxygen. As an example we have replaced He~{\sc ii} with 
 O~{\sc vii} in
the above model and find a good fit with $N_{\rm H}$ at the Galactic
value provided that the oxygen abundance in the IGM has almost the solar
value down to $z=3$ (setting $N_{\rm H}$ to the upper limit of
$\rm 1.8 \times 10^{20}\ cm^{-2}$ requires that the abundance of oxygen, all in
the form of O~{\sc vii}, must exceed 40 per cent of the solar value). The
abundance at least in oxygen must therefore be similar to that of the
intracluster medium at low redshifts.
This hypothesis can be tested by a search for the resonance
absorption lines of oxygen in distant quasars (see Aldcroft et al. 1994).

The most probable explanation is
that the downturn is due to absorbing
material intrinsic to the host galaxy of the
blazar.
This material needs to be metal-rich (or very high column
densities are required); our result of about $\rm 10^{22}\ cm^{-2}$
assumes solar metallicity. An interesting possibility is that this gas
is connected with the youth and possible formation of the host galaxy.
Large column densities may be reasonably expected in such young
objects (see e.g. Rees 1988; Fabian 1999). 
We note that the optical spectrum (Hook \& McMahon 1998), which covers
the rest-frame ultraviolet band of the object, indicates little dust
along our line of sight or any large  intrinsic absorption by hydrogen
(i.e. any very large damped Lyman $\alpha$ feature). This means that 
the absorbing material must be enriched but relatively dust free, and 
that the hydrogen in it must be ionized.

\section{Timing properties of GB~1428+4217}

Figure 5 (upper panel) displays the ROSAT PSPC light curve obtained for 
GB~1428+4217
between 1998 December 11 and 12.
A constant model fit 
to the data points can be rejected with 
$\rm >99$ per cent confidence. Similarly, if we work out the Poisson 
probability of obtaining the last two data points given the
mean of the first six data points, we obtain a probability of
$\le 2 \times 10^{-3}$. 

%\begin{figure*}
%\mbox{
%       \psfig{figure=gb1428_light_91.ps,width=5.7cm,clip=}
%       \psfig{figure=gb1428_light_92.ps,width=5.7cm,clip=}
%       \psfig{figure=gb1428_light_98.ps,width=5.7cm,clip=}}
%      \caption{
%ROSAT PSPC light curves of GB 1428+4217.
%The {\bf left panel} gives the light curve obtained during the
%ROSAT All-Sky Survey in January 1992.
%In the {\bf middle panel} the ROSAT light curve of GB 1428+4217 
%from an pointed observation centered on 4U 1417+42 in January 1993 is shown. 
%The light curve is corrected for vignetting effects and partial
%shadowing effects of the PSPC support rib (see Fabian et al. 1997).
%The {\bf right panel} gives the GB 1428+4217 ROSAT PSPC light curve
%obtained on December 11, 1998.
%Amplitude variability of about 50 per cent on a timescale of less then 
%one day is detected. As the variability scales with (1+z), the
%variations in the quasars frame are expected to about a factor of 3.
%}
%\end{figure*}

Using the mean of the first six data points and comparing it with the 
mean of the last two data points results to a 
variability amplitude of $25\pm 8$ per cent. 
No significant spectral variability is 
detected upon examination of appropriate hardness ratios.
Using the spectral parameters for GB~1428+4217 as displayed in Figure 2
the count rate variations translate into a 0.1--2.4~keV flux change in the
observer's frame of 
$\Delta F \rm = 5.9 \times 10^{-13}\ erg\ cm^{-2}\ s^{-1}$. 
The corresponding change in luminosity in the quasar's rest frame 
(0.6--13.7~keV) is 
$\Delta L \rm = 5.7 \times 10^{46} erg\ s^{-1}$ (using the best-fitting
spectral parameters and the cosmology given in Section 1)
within
$\Delta t \rm = 37000$~s, corresponding to about 6500~s in the rest frame.
Applying the
efficiency limit (Fabian 1979; Brandt et~al. 1999), we derive 
a remarkable efficiency of $\eta\ge 4.2$. 
The efficiency is
significantly larger than even that obtained for the narrow-line
quasar PHL 1092 (Brandt et~al. 1999).
Such a large derived efficiency strongly supports the evidence for 
relativistic flux boosting in the source and its blazar nature, 
previously discussed by Fabian et al. (1998). 

%The observed variability is remarkably strong for
%such a powerful source.

The mean 0.1--2.4 keV 
flux obtained from  the ROSAT observations of GB~1428+4217 between
1998 December 11 and 12 
is $F \rm = 2.7 \times 10^{-12}\ erg\ cm^{-2}\ s^{-1}$, 
corresponding to an isotropic luminosity in the quasar's rest frame 
of $L \rm = 2.6 
\times 10^{47} erg\ s^{-1}$.
The ROSAT PSPC flux value is somewhat larger than the mean flux values previously
reported for GB~1428+4217.
Fabian et al. (1997) give a mean 0.1--2.4 keV flux of 
$F \rm = 9.0 \times 10^{-13}\ erg\ cm^{-2}\ s^{-1}$ 
based on a ROSAT HRI observation on 1996 July 30.
The ASCA 2--10 keV flux obtained on 1997 January 17 translates into an
0.1--2.4 keV flux of $F \rm = 1.3 \times 10^{-12}\ erg\ cm^{-2}\ s^{-1}$
(Fabian et al. 1998).

\section{Summary}
In this paper we report on the detection of significant soft X-ray
absorption in the most distant X-ray detected object, the highly
X-ray luminous quasar GB~1428+4217, obtained during the last ROSAT observations
in December 1998. The low-energy sensitivity of the PSPC detector was employed to constrain
the intrinsic X-ray absorption. 
The soft X-ray
absorption of $N_{\rm H, fit}^{\rm z=0} \rm = (3.14 \pm 0.35) \times 10^{20}\ cm^{-2}$
is larger than the Galactic column of $N_{\rm H, Gal} \rm = (1.4 \pm 0.4) 
\times 10^{20}\ cm^{-2}$
towards GB~1428+4217 at the 5 sigma level.
The inferred intrinsic column density is
$N_{\rm H, fit}^{\rm z=4.72} \rm = (1.52 \pm 0.28) \times 10^{22}\ cm^{-2}$.
The most probable explanation for the soft X-ray absorption is intrinsic
absorption in GB~1428+4217. An interesting possibility is that this gas
is connected with the youth and possible formation of the host galaxy.
Absorption by intergalactic singly ionized helium (He~{\sc ii}), ionized
oxygen (O~{\sc vii})
or an intrinsic break in the
spectrum of GB~1428+4217 are found to be unlikely explanations for the soft X-ray
absorption. 
X-ray flux variability by $25\pm8$ per cent is detected on a time scale
of about 6500 s in the rest frame.  The derived efficiency limit of
$\eta\ge 4.2$ is remarkably large and further supports the blazar nature of
the object.

\section*{Acknowledgments}
We thank the anonymous referee for helpful and constructive suggestions
for further improvements of the paper. We thank J. Tr\"umper, P. Predehl and
D. Grupe for helpful discussions.
The ROSAT Project is supported by the Bundesministerium  f\"ur Bildung
und Forschung (BMBF/DLR) and the Max-Planck-Gesellschaft (MPG).
We gratefully acknowledge the support of the Royal Society (ACF) and 
NASA LTSA grant NAG5-8107 (WNB).

{}

\appendix
\section{Dependence of gain values on the X-ray absorption}
The nominal gain during the ROSAT PSPC observation of GB~1428+4217
is 102.7.
With Figure A1 we demonstrate that even in the case of unusual gain values 
the excess absorption above the Galactic column is still present. 

\begin{figure}
             \psfig{figure=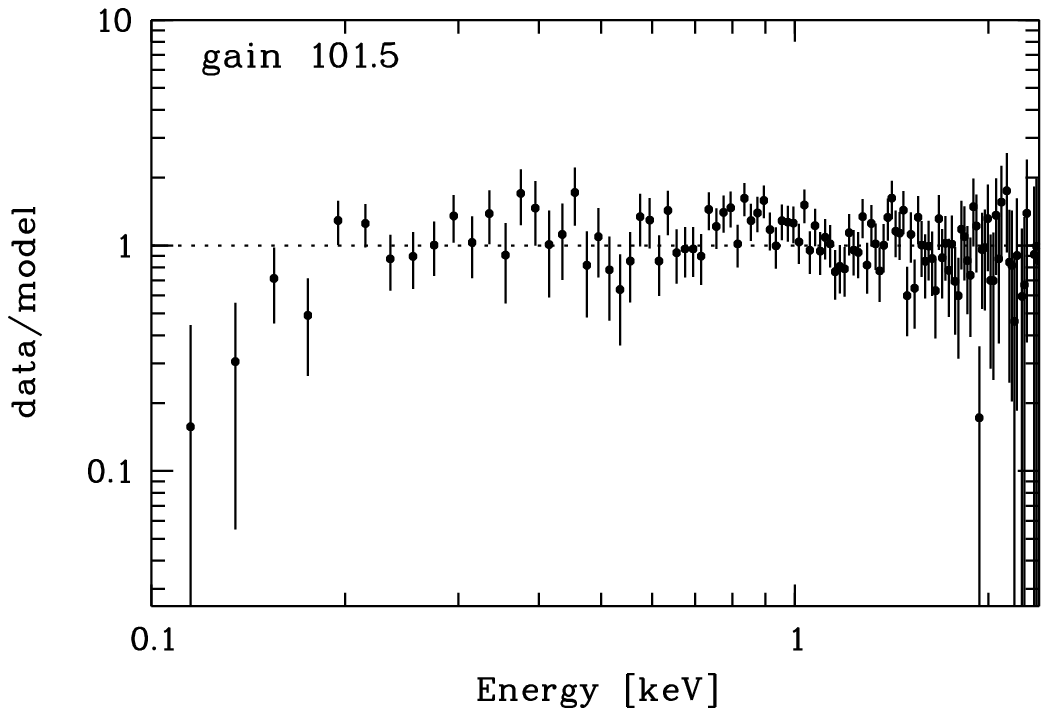,width=8.3cm,height=3.0cm,clip=}
             \psfig{figure=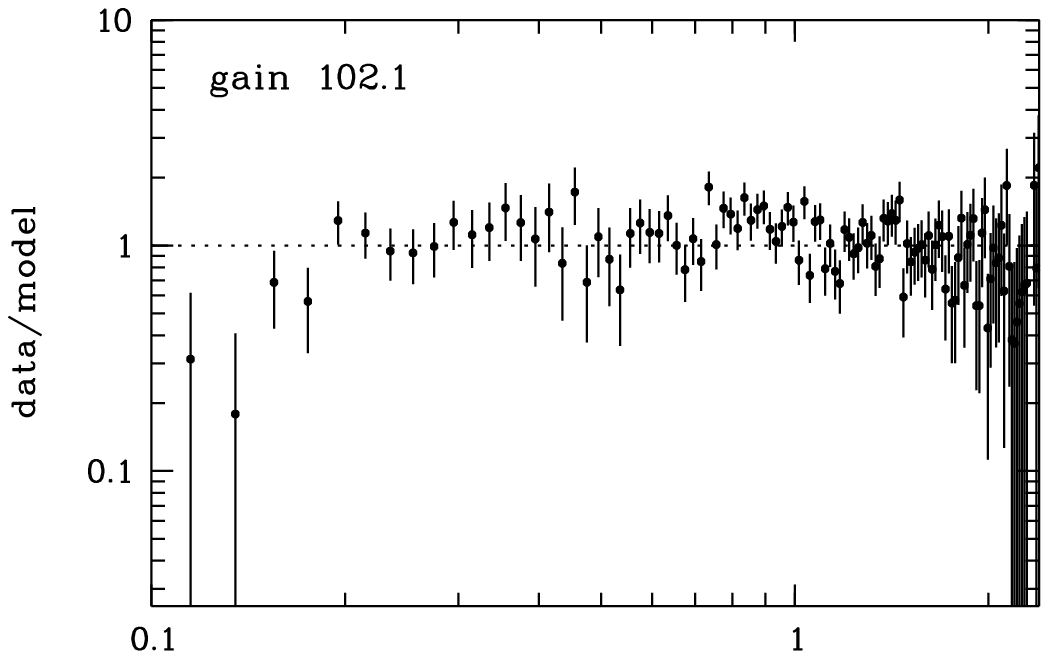,width=8.3cm,height=3.0cm,clip=}
             \psfig{figure=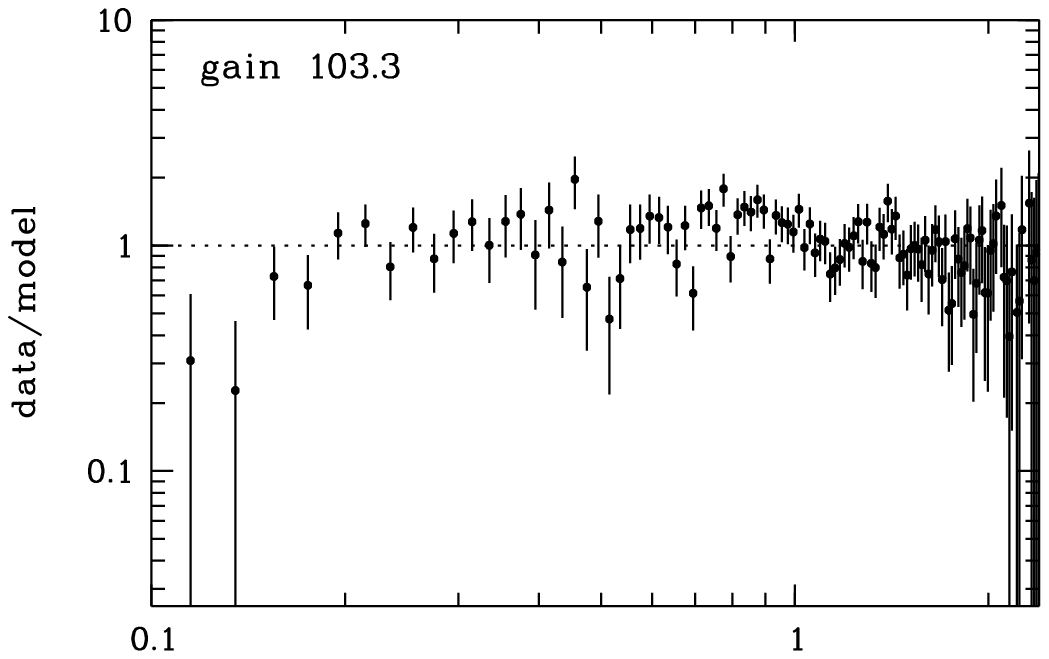,width=8.3cm,height=3.0cm,clip=}
             \psfig{figure=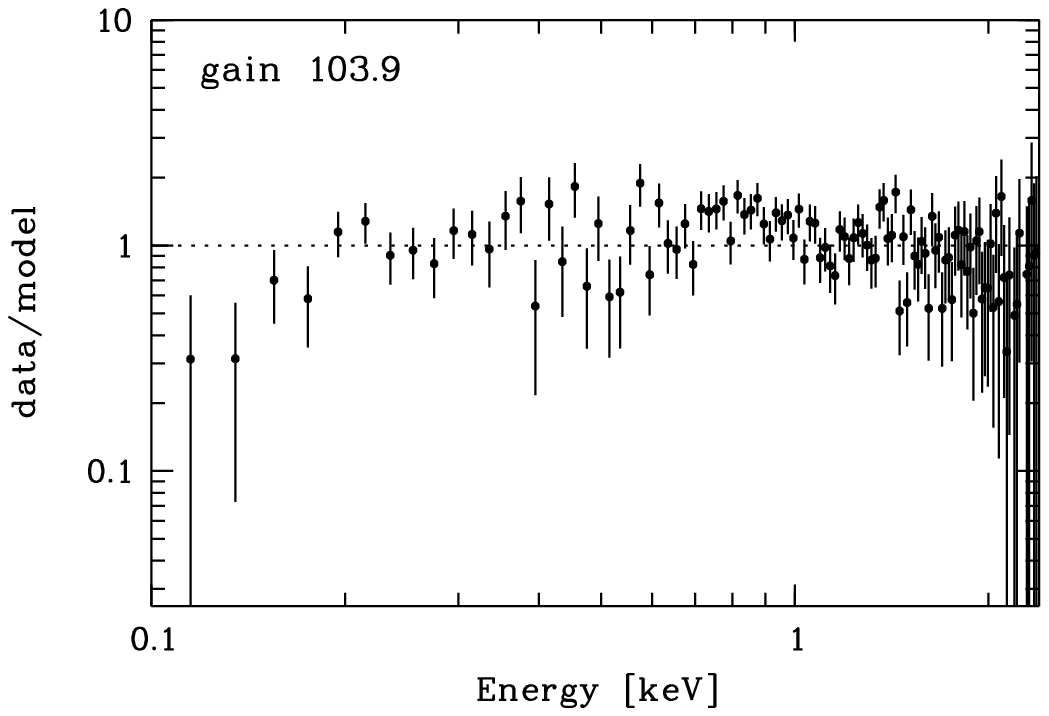,width=8.3cm,height=3.0cm,clip=}
      \caption{Influence of different gain corrections on the
soft X-ray absorption. We demonstrate that even in the case
of unusual gain corrections (101.5, 102.1, 103.3, and 103.9) the
excess absorption above the Galactic column density is still present.
A simple power-law fit
where the absorption column density and the photon index are
allowed to be free parameters is used to create the residuals.
}
\end{figure}

\end{document}